\documentstyle[11pt,newpasp,twoside,psfig]{article}
\newcommand{\kms}{km~s$^{-1}$}

\newcommand{\sL}{L$_{\odot}\,$}

\def\edcomment#1{\iffalse\marginpar{\raggedright\sl#1\/}\else\relax\fi}
\reversemarginpar

\begin{document}
\title{Halo streams as relicts from the formation of the Milky~Way}
 \author{Amina Helmi}
\affil{Leiden Observatory, P.O. Box 9513, 2300 RA Leiden, The Netherlands}

\begin{abstract}
In this contribution we discuss the expected properties of our
halo if our Galaxy had been built from mergers of smaller
systems. Using both analytic arguments and high-resolution
simulations, we find that the Galactic halo in the Solar neighbourhood
should have a smooth spatial distribution and that its 
stellar velocity ellipsoid should consist of several hundred kinematically
cold streams, relicts from the different merging events.  We also
discuss observational evidence which supports this hierarchical
picture: two stellar streams originating in the same system that
probably fell onto the Milky Way about ten Gyr ago. Finally we address
what future astrometric missions like FAME or GAIA may reveal by
observing the motions of nearby halo stars.
\end{abstract}

\section{Introduction}

In recent years, considerable theoretical effort has been put into
understanding the properties of galaxies within the hierarchical
paradigm of structure formation in the Universe (e.g. Kauffmann et al.
1999).  Our own Galaxy can play a very important role in constraining
these models. For the Milky Way we have access to kinematics,
distances, ages and chemical compositions of individual stars,
information which is available for no other system. Thus observable
substructure leftover by the different merging events could be just at hand.

The natural place to look for the stars of the various systems that merged
to build up the Milky Way is the Galactic halo, since as such systems
disrupt they leave trails of stars along their orbits. An ensemble of
disrupted galaxies would thus naturally produce a spheroidal
component.  What imprints do these mergers leave on the present day
distribution of halo stars?  What are the observational requirements
(accuracies, sample sizes) that will enable us to test from this
perspective if our Galaxy was built hierarchically?

\section{Characterising the signatures: Theoretical approach}

Numerical simulations of the disruption of satellite galaxies which
probe the outer parts of the Galaxy {\it only}, show that, after many Gyr of
evolution, their stars are distributed in coherent spatial structures
almost along great circles in the sky (Johnston, Hernquist \& Bolte
1996).  When the satellite's orbits probe the inner Galaxy (i.e. with
apocentres smaller than 30--40 kpc), the Galactic disk (breaking the
spherical symmetry) makes the orbital plane precess in space thus
transforming the two-dimensional streams in three-dimensional
structures. The stars appear now distributed almost homogeneously in
space, and so the signatures of the different mergers experienced have
disappeared (almost completely) in configuration space as shown in Figure~1.
\begin{figure}
\hspace*{2.5cm}\psfig{figure=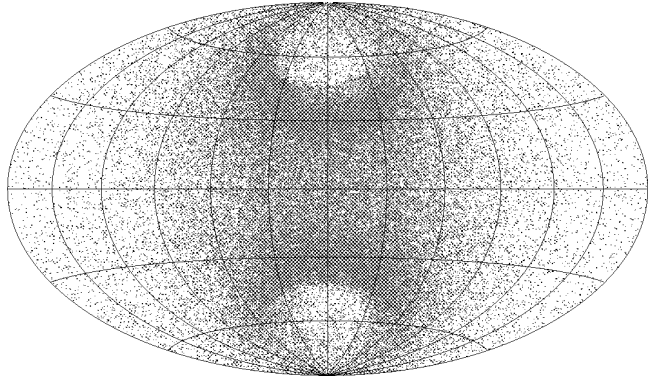,height=5cm}
\caption{Sky distribution (in Galactic ($l$,$b$) coordinates) of the debris of a satellite galaxy after 10
Gyr of evolution. In this simulation, the Milky Way was represented by
a fixed potential with three components: halo, disk and bulge.  The
satellite was on an orbit with an apocentre $\sim$16 kpc and a
pericentre of $\sim$7 kpc. The initial core radius of the satellite
was $0.5$ kpc and its initial velocity dispersion 16.5~\kms.}\label{fig:sky_sim}
\end{figure}

Because phase-space density is a conserved quantity, the satellite galaxies 
initial high values strongly constrain how the debris can
spread out in phase-space. As the satellite's stars spread out in
space, they define streams: kinematic structures such as those shown
in Figure~2.  These streams are the characteristic
signatures of mergers, and should thus be visible in the kinematics of
halo stars in the Solar neighbourhood if our Galaxy was indeed built
hierarchically.
\begin{figure}[b]
\psfig{figure=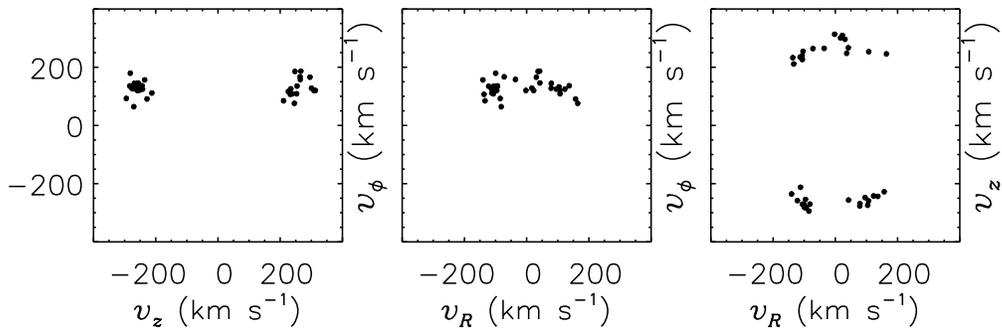,height=4.3cm}
\caption{Scatter plots of the velocity components for particles located
inside boxes of 1 kpc on a side on the ``Solar circle'' for the  
simulation described in Figure~1.}\label{fig:vel_sim}
\end{figure}

Helmi \& White (1999) find that, if the whole $10^9$ \sL stellar halo
was built from disrupted satellites, the velocity ellipsoid in the
Solar neighbourhood should consist of 300 -- 500 stellar streams with
internal velocity dispersions of $\sim$ 5 \kms. These results are
based on numerical simulations and analytic calculations that model
the disruption of a single satellite galaxy, and the evolution of its
debris streams in a static potential.  But note that this case is
dramatically different from the hierarchical picture that we hope to
test here. 

With  this in  mind, we  analysed high-resolution  simulations  of the
formation  of   a  cluster  in  a   $\Lambda$CDM  cosmology  (Springel
1999). After  scaling down the cluster  to a galactic size  halo (by a
factor $\sim$ 12  in radius and in circular  velocity) we analysed the
evolution of the debris originating in the different halos that merged
to build up the ``Galaxy''. We found that the debris streams behave in
a similar fashion to the  streams moving in a static potential (Helmi,
White  \&  Springel,  in  preparation).   From  these  high-resolution
simulations we  can determine the  number of dark-matter streams  as a
function of distance from the halo  centre. We find that in what would
be the ``Solar  neighbourhood'' of the scaled halo,  one should expect
approximately 1000 {\it dark} streams (Figure~3).  Note, however, that
the  scaling used above  is independent  of the  assembly time  of the
halo, which thus means that one may expect a slightly larger number of
streams  for a  galactic  halo than  for  a scaled-down  version of  a
cluster halo.   In general a  galactic halo (being assembled  about 10
Gyr ago) will have had more time to relax than  a cluster halo (typically
assembled $\sim$ 8 Gyr ago), and so the streams to spread out
in space.  We  estimate that this number cannot be  more than a factor
$\sim 2$ larger.
\begin{figure}[!b]\vspace*{-.5cm}
\hspace*{2.5cm}\psfig{figure=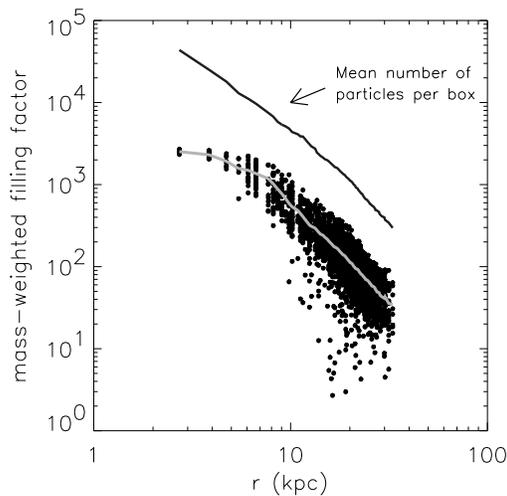,height=6.5cm}
\caption{Mass-weighted number of streams in boxes of 2.72 kpc on a side
for a high-resolution simulation (66 million particles) of the
formation of a cluster in a $\Lambda$CDM cosmology scaled to a Milky
Way halo, as a function of distance from the centre of the scaled
halo. The thick grey line going through the data points corresponds to
the median mass-weighted number of streams, derived for boxes at the
same distance from the centre.}\label{fig:ff}
\end{figure}

How do these results compare to our estimates derived for the stellar
halo before? How many {\it star} streams should we observe in the
Solar neighbourhood? Each dark matter halo will contribute by about a
factor of ten more dark streams than stellar ones, since baryons are
generally clustered in the inner roughly one-tenth of their host
halo. Also due to this segregation, the baryons orbital properties do
not exactly follow those of their dark matter counterparts. A typical
star in the Solar neighbourhood has a period of 0.25 Gyr, with an
apocentre of about 11 kpc. A typical dark matter particle will have an
apocentre of the order of 20 kpc, corresponding to an orbital
timescale of the order of 0.3 Gyr.  Because the time for dispersal
scales as $1/t^3$, this implies that the naively expected number of
star streams should be doubled. In synthesis, if we break the 2000
dark streams (now scaled to take into account the earlier assembly of
a galactic halo) into ten contributors, each one of these will give
rise to $20 \times 2$ stellar streams, if our scalings are roughly
right. This estimate brings us close to the result derived initially:
we should expect about four hundred stellar streams in the Solar
neighbourhood.

\section{Finding the streams: Observational approach}

Over the years, an increasing number of observations have been
suggesting the presence of substructure in the halo of the Galaxy
(e.g. Eggen 1965; Rodgers \& Paltoglou 1984; Majewski, Munn \& Hawley
1994).  Detections of lumpiness in the velocity distribution of halo
stars are becoming increasingly convincing, and the discovery
of the Sagittarius dwarf satellite galaxy (Ibata, Gilmore \& Irwin
1994) is a dramatic confirmation that accretion and merging continue
to affect the Galaxy. However {\it direct} evidence that the bulk of
the Milky Way's population of old stars was built up from mergers had
been lacking until quite recently. Using kinematic data from the
HIPPARCOS satellite, Helmi et al. (1999) have found two halo star
streams which share a common progenitor: a single coherent object
disrupted during or soon after the Milky Way's formation, and which
probably resembled the Fornax and Sagittarius dwarf spheroidal
galaxies.  The kinematic properties of the sample of halo stars used
are shown in Figure~4, with the stars identified as
members of the streams highlighted. This figure should be compared
to Figs.~1 and 2, which correspond to a simulation set up to reproduce
the properties of the streams.  The presence of these streams was also
confirmed later by Chiba \& Beers (2000) in a much larger sample.
\begin{figure}
\psfig{figure=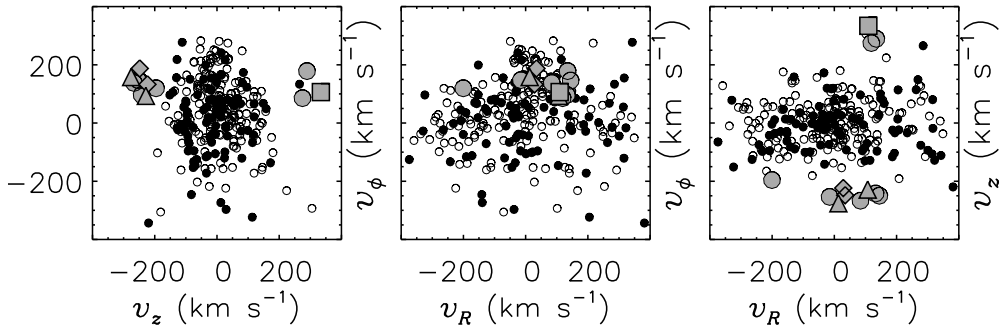,height=4.3cm}
\caption{Distribution of nearby halo stars in velocity space for our
original sample of giants with [Fe/H] $\le -1.6$ dex and distances
$\le $ 1 kpc (solid circles) and for the extended sample of more
metal-rich and more distant giants (open circles).  Candidates for our
detected substructure are highlighted in grey: triangles indicate more
metal--rich giant stars at distances $> 1$~kpc, diamonds more
metal--rich giants at $\le 1$~kpc, squares metal--poor giants at
$>1$~kpc, and circles metal-poor giants at $\le 1$~kpc. Here we used
kinematic data from the HIPPARCOS satellite, and radial velocities,
metal abundances and photometric distances obtained from ground based
data.}\label{fig:vel_HIPP}
\end{figure}

Samples of halo stars in the vicinity of the Sun are currently too
small to be useful to fully test the hierarchical picture and the
predictions made in the previous section. Since the expected number of
streams is approximately 500, the required samples should preferably
have 2000 or more stars with good kinematics. To detect the clumpy
nature of the velocity ellipsoid in the Solar neighbourhood, this
ellipsoid should be broken up into the streams, in which case the
required accuracy should be $\sigma \le 120/(500)^{1/3}$ \kms or
$\sigma \le 15$ \kms, values that can be reached presently. To detect
the individual streams one may require a 3$\sigma$ distinction or 5
\kms accuracy.

\section{Testing the paradigm: The future}

Future astrometric satellites will measure with very high accuracy the
motions of thousands to many millions of stars in our Galaxy. Whereas
SIM is a targeted mission (and will thus not be useful to test in the
Solar neighbourhood the predictions described above), FAME promises to
measure positions and parallaxes for about $4 \times 10^7$ stars. For
stars brighter than $V \sim 5$ it will do so to better than 50 $\mu$as
in parallax and 50 $\mu$as yr$^{-1}$ in proper motion, and at $V\sim 15$ these
accuracies will be degraded by an order of magnitude.  On the other
hand, the proposed ESA astrometric satellite GAIA will provide very
precise astrometry ($<$10 $\mu$as in parallax and $<$10 $\mu$as
yr$^{-1}$ in proper motion at $V \sim 15$, increasing to 0.2 mas
yr$^{-1}$ at $V\sim 20$) and multicolour photometry, for all 1.3
billion objects to $V\sim20$, {\it and} radial velocities with
accuracies of a few \kms~ for most stars brighter than $V\sim17$, so
that full and homogeneous six-dimensional phase-space information will
be available.

Helmi \& de Zeeuw (2000) have run numerical simulations of the
disruption of satellite galaxies in a Galactic potential with the aim
of building up the entire stellar halo. Using these simulations, they
generate artificial FAME and GAIA catalogues in which they look for
the signatures left by the accreted satellites. They find that
disrupted satellite galaxies may be recovered after a Hubble time as
lumps in the space of integrals of motion. Using energy, angular
momentum and its $z$-component to define such a space, they look for
the clumps using a Friends-of-Friends algorithm. For a simulated GAIA
catalogue they are able to recover 50\% of all accreted objects,
whereas for a FAME catalogue (using radial velocities accurate to 15
\kms) the recovery rate is $\sim 15$\%, an effect which is due
to the lower accuracies and the brighter limiting magnitude (i.e. the
samples with good kinematic information are smaller).  However both
missions should be able to test the hierarchical formation paradigm on
our Galaxy by measuring the amount of halo substructure in the form of
nearby kinematically cold streams. In Figure~5 the two-point
correlation function in velocity space is shown for four realizations
of the artificial FAME and GAIA catalogs, each corresponding to a
different position of the ``Sun'' along the ``Solar circle''. The
presence of kinematically cold streams is visible as an excess of
pairs in the bins corresponding to small velocity differences. And,
even with velocity errors of the order of 20 \kms, it will be possible
to determine that the clumpy nature of the halo velocity ellipsoid in
the Solar neighbourhood.
\begin{figure}
\psfig{figure=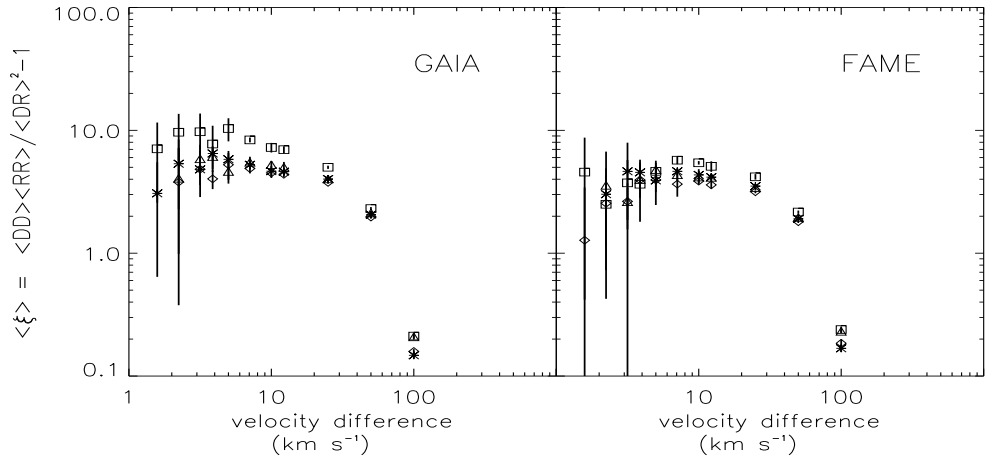,height=6cm,width=13cm,angle=90}
\caption{The two-point correlation function for particles (assuming
luminosities of giant stars) inside spheres of 1 kpc radius around the
Sun (defined as 8 kpc from the Galactic centre on the Galactic disk)
computed as the weighted average over five realizations of the FAME
and GAIA catalogues. The different symbols correspond to spheres at
different locations of the ``Sun'' on the Solar
circle.}
\end{figure}

\vspace*{0.15cm}
\noindent {\bf Acknowledgments.} I wish to thank my collaborators Tim de Zeeuw, Volker Springel and Simon White.

\end{document}